\documentclass[aps,pra,twocolumn,amsmath,amssymb,showpacs,superscriptaddress]{revtex4}

\usepackage[dvips]{graphicx}
\usepackage{mathrsfs}
\input{Qcircuit}

\begin{document}

\title{Error propagation in loss- and failure-tolerant quantum computation schemes}

\author{Peter P. Rohde}
\email[]{rohde@physics.uq.edu.au}
\homepage{http://www.physics.uq.edu.au/people/rohde/}
\affiliation{Centre for Quantum Computer Technology, Department of Physics\\ University of Queensland, Brisbane, QLD 4072, Australia}

\author{Timothy C. Ralph}
\affiliation{Centre for Quantum Computer Technology, Department of Physics\\ University of Queensland, Brisbane, QLD 4072, Australia}

\author{William J. Munro}
\affiliation{Hewlett-Packard Laboratories, Filton Road, Stoke Gifford, Bristol BS34 8QZ, UK}
\date{\today}

\frenchspacing

% ABSTRACT
\begin{abstract}
Qubit loss and gate failure are significant obstacles for the implementation of scalable quantum computation. Recently there have been several proposals for overcoming these problems, including schemes based on parity and cluster states. While effective at dealing with loss and gate failure, these schemes typically lead to a blow-out in effective depolarizing noise rates. In this supplementary paper we present a detailed analysis of this problem and techniques for minimizing it.
\end{abstract}

\pacs{03.67.Lx,42.50.-p}

\maketitle

% INTRODUCTION
\section{Introduction}
Quantum computation has the potential to solve computational problems intractable on classical computers \cite{bib:NielsenChuang00}. A major obstacle facing the experimental realization of quantum computing is the introduction of errors. In some architectures the dominant errors are qubit loss and gate failure. This is especially the case for photonic schemes, such as linear optics quantum computing \cite{bib:KLM01}, where these effects arise due to physical loss of photons, source and detector inefficiencies, and the inherent non-determinism of multi-qubit gates. This has motivated the development of recent schemes for tolerating qubit loss \cite{bib:RalphHayes05,bib:Varnava05} and gate failure \cite{bib:Duan05,bib:BarrettKok05}.

These schemes achieve loss/failure tolerance through the introduction of redundant encoding. This provides multiple attempts to perform the relevant operations, suppressing loss/failure rates. However, redundancy also introduces additional opportunities for noise to be introduced, increasing effective error rates. In general this results in an exponential blow-out in depolarizing noise \cite{bib:RohdeRalphMunro07}. This problem of error blow-out can be minimized by modifying the schemes. However, in general this significantly reduces their loss/failure tolerance.

In a variety of contexts this is a serious problem. When embedded into a fault tolerant quantum computing architecture it significantly reduces the effective fault tolerant threshold. In a loss-tolerant quantum memory scenario it quickly reduces the memory to a depolarizing (i.e. classical) channel. In the context of state preparation strategies, which have applications beyond quantum computing, it results in the preparation of highly mixed states.

We demonstrate this principle by example of three well known protocols for tolerating qubit loss and gate failure, and expand on the work of Ref. \cite{bib:RohdeRalphMunro07} by deriving analytic expressions for effective error rates. We also examine techniques for minimizing the error scaling problem. We begin with a general discussion of the relevant error propagation properties of cluster (or graph) states in section \ref{sec:graph_state_disc}. Next we discuss a gate-failure tolerant approach to preparing cluster states in section \ref{sec:failure}. In principle this scheme can tolerate arbitrary non-zero gate success probabilities, at the expense of a polynomial overhead in physical resource requirements. We demonstrate that effective error rates increase exponentially with gate failure probability, which places practical limitations on the failure tolerance of this scheme. We present a more general state preparation strategy, which allows us to reduce this effect at the expense of an overhead in physical and temporal resource requirements. In section~\ref{sec:cluster} we consider a recent scheme for tolerating qubit loss in the cluster state model for quantum computation. In principle this scheme can tolerate extremely high loss rates -- up to 50\%. We demonstrate that, like the previous scheme, this protocol has the effect of magnifying the effects of depolarizing noise, which places practical limitations on realistically tolerable loss rates. We discuss a method for minimizing this effect, relying on majority voting techniques. However, doing so significantly reduces the loss tolerance of the scheme. Again we derive expressions for effective error and loss rates. Finally, in section \ref{sec:parity} we consider the loss-tolerant parity scheme. As before, this scheme is shown to magnify depolarizing noise. We discuss a potential solution to this, also based on a majority voting technique. We conclude in section \ref{sec:conclusion}.

Our results indicate that in realistic scenarios the loss/failure tolerance of specialized schemes is likely to be significantly lower than is possible in principle. Consequently, such scheme are likely to be limited to dealing with comparatively modest levels of loss and failure.

% DISCUSSION OF GRAPH STATE PROPERTIES
\section{Discussion of graph state properties} \label{sec:graph_state_disc}
We begin with a general discussion of the behavior of graph states (in this paper we use the terms `graph state' and `cluster state' interchangeably) under qubit loss and single qubit errors. We assume familiarity with the cluster state model for quantum computation and suggest the unfamiliar reader refer to Refs.~\cite{bib:Raussendorf01,bib:Raussendorf03,bib:Nielsen06}. In general such behavior is highly topology-dependent. We focus on two topologies of particular interest -- linear graphs, and fully connected graphs (which are locally equivalent to the maximally entangled GHZ states).

\subsection{The stabilizer representation of graph states}
An $n$ qubit graph state can be expressed in terms of a set of $n$ stabilizers, one for each vertex in the graph. The stabilizers take the form
\begin{equation}
\hat{S}_i = \hat{X}_i \bigotimes_{j\in v(i)} \hat{Z}_j,
\end{equation}
where $i$ denotes a qubit, $v(i)$ denotes the neighborhood of $i$, and $\hat{X}_i$ and $\hat{Z}_i$ denote the usual Pauli bit-flip and phase-flip operators respectively, acting on qubit $i$.

The stabilizer representation is particularly well suited to understanding the propagation of errors through graph states, which we consider in the following subsections.

\subsection{Recoverability from qubit loss}
In general, a graph state which has suffered the loss of a qubit (or alternately any form of located error) can be partially recovered by measuring the qubits adjacent to the lost qubit in the $Z$-eigenbasis. This feature of graph states enables, for example, the gate-failure tolerant construction of Refs.~\cite{bib:Duan05,bib:BarrettKok05} that we discuss in the next section.

Because this recovery operation requires measuring the neighborhood of the affected qubit, it is highly topology-dependent. Consider a linear graph of length $n$. When the $m^\mathrm{th}$ qubit is lost one can recover two linear graphs of length $m-2$ and $n-m-2$ by measuring the qubits adjacent to the lost qubit in the $Z$-eigenbasis. For a connected graph $K_n$ (i.e. a GHZ state) the behavior is very different. Because the graph is fully connected, the recovery operation requires measuring every remaining qubit, which completely destroys the system. Thus, for linear graphs, under qubit loss much of the entanglement is preserved, whereas for GHZ states it is completely destroyed.

\subsection{Error propagation under $Z$-measurement}
We now consider the situation where we perform a sequence of $Z$-measurements on a graph state. This situation arises naturally in many graph state preparation strategies where unwanted qubits are removed from the graph via $Z$-measurement. In general, an error in the $Z$-measurement outcome (i.e. an $X$-error) will propagate $Z$-errors onto the neighbors of the affected qubit. This property follows directly from the stabilizer represenation. Suppose we have a graph state $\ket\psi$ which is subject to an $\hat{X}$ error at location $i$,
\begin{equation}
\hat{X}_i\ket\psi = \hat{X}_i\hat{S}_i\ket\psi = \bigotimes_{j\in v(i)} \hat{Z}_j \ket\psi.
\end{equation}

Consider a linear graph where we measure a qubit in the $Z$-eigenbasis. This has the effect of dividing the graph into two smaller linear graphs. If an $X$-error was present on the measured qubit, $Z$-errors will propagate onto the end qubits of each of the newly created linear graphs. Importantly, if a sequence of $Z$-measurements is performed along a linear graph segment, only errors affecting the terminating qubits will propagate onto the remaining state. Thus, there is no accumulation of errors as we perform the measurement sequence.

For the fully connected graph $K_n$, $Z$-measurement of a single qubit reduces the graph to $K_{n-1}$ and propagates a correlated $Z$-error onto all of the remaining qubits (i.e. $Z^{\otimes n-1}$. It can easily be verified from the stabilizer representation that this is equivalent to a $Y$-error acting on any one qubit in the remaining state. If we again consider the situation where we perform a sequence of $Z$-measurements, unlike the linear graphs we now have a situation where the propagated $Y$-errors accumulate. Specifically, the net probability of a $Y$-error being propagated onto the remaining state is the probability of an odd number of measurement errors occurring. This exhibits exponential dependence on the number of measured qubits.

\subsection{Error propagation under $X$-measurement} \label{sec:X_prop_rules}
Next we consider the situation where we perform $X$-measurements on qubits in a graph state. First consider the linear graph. We can quickly establish the error propagation properties from the following two circuit identities. The first states that a $Z$-error acting on a qubit which is subsequently measured in the $X$-eigenbasis is equivalent to an $X$-error on a neighboring qubit. In the circuit model this can be represented as,
\begin{displaymath}
\Qcircuit @C=1em @R=.7em{
& \ket{+} & & \ctrl{1}  & \gate{Z} & \measure{X} & = & & \ket{+} & & \ctrl{1}  & \qw      & \measure{X} \\
& \ket{+} & & \ctrl{-1} & \qw      & \qw         &   & & \ket{+} & & \ctrl{-1} & \gate{X} & \qw         \\}
\end{displaymath}
The second identity states that when a $Z$-error is introduced onto a graph state qubit, and both that qubit and its neighbor are measured in the $X$-eigenbasis, this is equivalent to a $Z$-error acting on the third qubit along the chain,
\begin{displaymath}
\Qcircuit @C=1em @R=.7em{
& \ket{+} & & \ctrl{1}  & \qw       & \gate{Z} & \measure{X} &   & & \ket{+} & & \ctrl{1}  & \qw       & \qw      & \measure{X} \\
& \ket{+} & & \ctrl{-1} & \ctrl{1}  & \qw      & \measure{X} & = & & \ket{+} & & \ctrl{-1} & \ctrl{1}  & \qw      & \measure{X} \\
& \ket{+} & & \qw       & \ctrl{-1} & \qw      & \qw         &   & & \ket{+} & & \qw       & \ctrl{-1} & \gate{Z} & \qw}
\end{displaymath}
Suppose we have the situation where we wish to sequentially measure a series of qubits from a linear graph in the $X$-eigenbasis. We are interested in how errors propagate onto the terminating qubit, which we refer to as the `root' qubit. From the above two circuit identities we can establish the following. When a $Z$-error is introduced onto qubits which are an even number of qubits away from the root qubit, after measurement this is equivalent to a $Z$-error acting on the root qubit. Secondly, when a $Z$-error is introduced onto a qubit an odd number of qubits away from the root qubit, after measurement this is equivalent to an $X$-error acting on the root qubit. In both of these cases the propagated errors accumulate and the net error probability is the probability of an odd number of the respective errors being propagated - i.e. exponential dependence.

For the fully connected graph the situation is similar to before. Namely, an $X$-measurement on $K_n$ reduces it to $K_{n-1}$, and an incorrect measurement result propagates a $Y$-error onto one of the remaining qubits. As before, if a sequence of measurements is performed this leads to an accumulation of errors with exponential dependence on the number of measured qubits.

% SCALABLE PROBABILISTIC QUANTUM COMPUTING AND STATE PREPARATION SCHEMES
\section{Scalable probabilistic quantum computing and state preparation schemes} \label{sec:failure}
In some quantum computing architectures gate failure is a significant problem. This undermines our ability to perform scalable quantum computing, because the probability of a computation succeeding drops exponentially with the size of the circuit. Recently schemes have been suggested for tolerating gate failure \cite{bib:Duan05,bib:BarrettKok05}. We specifically consider the scheme of Ref.~\cite{bib:Duan05}, which describes a technique for constructing cluster states, a resource for universal quantum computation, using physical resources that grow polynomially with the size of the desired cluster, thus allowing for `efficient' quantum computation. In principle this scheme can tolerate arbitrary non-zero gate success probabilities.

The scheme is an example of a `divide-and-conquer' approach to state preparation. A related proposal is Nielsen's \cite{bib:Nielsen04} \emph{micro-cluster} approach. We now briefly review this scheme. Ordinarily, using deterministic gates, if we wish to prepare, say, a square lattice cluster, we begin with a lattice of qubits initally prepared in the $\ket{+}=(\ket{0}+\ket{1})/\sqrt{2}$ state and perform a sequence of {\sc CPHASE} gates between nearest neighbors. Using non-deterministic gates this is clearly not possible since the success probability is exponentially small.

\begin{figure}[!htb]
\includegraphics[width=\columnwidth]{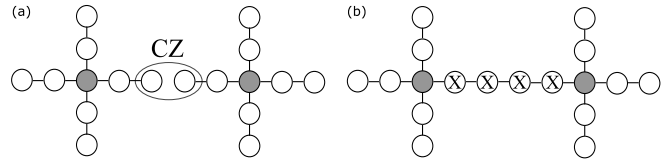}
\caption{Gate-failure-tolerant approach to constructing cluster states. The fundamental building block is the `+'-cluster. This has a central node (shown in gray) which will ultimately belong to the constructed square lattice. The central node is bonded to four linear chain clusters, each of length $n_l$. These `arms' provide redundancy, allowing multiple bonding attempts. To grow a cluster, rather than bond two cluster qubits together directly, we utilize +-clusters and attempt to bond them starting at the ends of the arms (a). Whenever this fails we lose two qubits from the respective arms, but can recover the remainder of the cluster by measuring the neighboring qubits in the $Z$-eigenbasis. We can keep reattempting the gate until there are no qubits remaining in the arms. When bonding succeeds we have the two desired cluster nodes with some remaining arm qubits left between them. These are removed by measuring them in the $X$-eigenbasis (b).} \label{fig:failure}
\end{figure}

This scheme overcomes this problem by utilizing a resource of `+'-clusters. These consist of a central node, which will ultimately belong to the prepared cluster (see Fig.~\ref{fig:failure}), from which emanate four linear chain clusters of length $n_l$. Using non-deterministic gates these states can be prepared off-line in advance. Utilizing this resource we proceed to bond two +-clusters together by performing a {\sc CPHASE} gate between the ends of their arms, as shown in Fig.~\ref{fig:failure}a. If this fails we can recover the remainder of the +-clusters by performing $Z$-measurements on the qubits neighboring the ones to which we applied the {\sc CPHASE} gate. Thus, a gate failure reduces arm length by two qubits. We can now reattempt this bonding using what is left of the arms. Thus, an arm length of $n_l$ allows for $n_l/2$ bonding attempts. When bonding finally succeeds (assuming it does) any left-over arm qubits may be removed by measuring them in the $X$-eigenbasis, as shown in Fig.~\ref{fig:failure}b. This leaves us with a cluster state where the two central qubits from the original +-clusters are now neighbors.

It is clear that this structure provide redundancy in the bonding process, giving us multiple attempts at bonding qubits together, thereby suppressing the effective gate failure rate. It is shown in Ref.~\cite{bib:Duan05} that this procedure generalizes to the construction of arbitrarily large square lattice clusters. Furthermore, it is shown that the required arm length of the resource of +-clusters scales as,
\begin{equation} \label{eq:arm_len_scale}
n_l\approx \frac{2}{p_g}\mathrm{ln}\left(\frac{2N}{\varepsilon}\right),
\end{equation}
where $p_g$ is the success probability of the {\sc CPHASE} gate, $N$ is the desired number of qubits in the final cluster state, and $\varepsilon$ the probability of successfully preparing the cluster state.

\subsection{Analysis of error propagation}
This scheme relies on sequential $X$-measurements along a linear graph segment to remove left-over qubits. Thus we can use the error propagation rules from section \ref{sec:X_prop_rules}. From these rules we know that a $Z$-error will be propagated onto the root qubit if an odd number of $Z$-errors occurred on half of the measured qubits, and similarly for $X$-errors. Thus, the effective error rates can be expressed,
\begin{eqnarray}
p_X'&=&\sum_{\substack{0\leq i\leq n_X/2 \\ i\in 2\mathbb{Z}+1}}\binom{n_X/2}{i}{p_X}^i(1-p_X)^{n_X/2-i}\nonumber\\
p_Z'&=&\sum_{\substack{0\leq i\leq n_X/2 \\ i\in 2\mathbb{Z}+1}}\binom{n_X/2}{i}{p_Z}^i(1-p_Z)^{n_X/2-i},
\end{eqnarray}
where $p_X$ and $p_Z$ are the physical $X$ and $Z$-error rates, and $n_X$ is the number of measured qubits. %For $p_e=p_X=p_Z$ the effective error rate is approximately,
%\begin{equation}
%p_\mathrm{error}(n_X)\approx 2\sum_{\substack{0\leq i\leq l \\ i\in 2\mathbb{Z}+1}}\binom{n_X/2}{i}{p_e}^i(1-p_e)^{n_X-i}.
%\end{equation}
Thus, the effective $X$- and $Z$-error rates exhibit exponential dependence on $n_X$, which, from Eq.~\ref{eq:arm_len_scale}, is inversely proportional to gate success probability. Thus, while this scheme can tolerate arbitrary non-zero gate success probabilities in principle, in practise noise blow-out places a practical limitation on gate success probability.

\subsection{Minimizing the effects of error propagation}
We now discuss a technique for minimizing the effects of error blow-out in divide-and-conquer based state preparation strategies. The approach is effectively to trade an increase in physical resource requirements for a reduction in accumulated error rates. This is achieved by beginning with larger resource states, which are prepared using a `single-shot' approach, i.e. probabilistically prepared in one attempt.

\begin{figure}[!htb]
\includegraphics[width=\columnwidth]{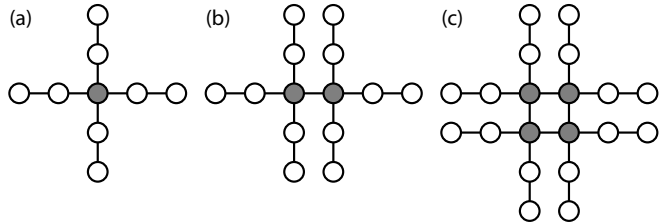}
\caption{Different resource states that can be employed in the scalable construction of cluster states using non-deterministic gates.} \label{fig:resource}
\end{figure}

Consider the Duan \emph{et al.} scheme. Referring to Fig.~\ref{fig:resource}, we begin with a resource of clusters of the form shown in (a). We then fuse two such clusters together and measure out the redundant qubits to produce a resource of clusters of form (b). Similarly, fusing two of these clusters together and removing the redundant qubits yields a cluster of form (c). Suppose the initial resource of +-clusters are produced using a single-shot approach. Thus, we assume the initial resource states do not suffer from accumulated errors. Ordinarily a (b)-type cluster suffers error accumulation associated with the measurement of redundant qubits from two fused +-clusters. This can obviously be avoided by instead beginning with a resource of (b) clusters, directly prepared using a single-shot approach. Doing so avoids the measurement of the interstitial redundant qubits and the associated accumulation of errors. Obviously this idea can be used to an arbitrary extent, allowing for further suppression of error accumulation effects. However, the degree to which this approach can be employed is practically limited by the exponential scaling of single-shot preparation.

This technique effectively allows us to tailor a strategy which presents an arbitrary tradeoff between the single-shot and divide-and-conquer strategies, where the level of tradeoff is limited by the gate failure rate. The tradeoff between competing resources is clear. For a given bound on the effective error rate, using larger resource states allows us to tolerate higher local error rates, since error accumulation is reduced. However, because the resource states are prepared using a single-shot approach, preparing larger resource states requires physical and temporal resources growing exponentially with the size of the resource state, and polynomially with gate failure probability. This places fundamental limitations on practically tolerable gate failure rates.

%Let us consider a specific example in detail. Suppose we begin with a resource of +-clusters that we wish to join together to form a resource of (b)-type clusters. The average number of redundant qubits (per final cluster qubit) that must be measured out following successful bonding of the arms is given by
%\begin{equation}
%\langle m\rangle = p_\mathrm{gate}n_l + \sum_{f=1}^{n_l/2} (n_l - 2f + 1) p_\mathrm{gate}(1-p_\mathrm{gate})^f.
%\end{equation}
%Here we are summing over the number of gate failures that occur before a successful bonding occurs. The first term corresponds to an immediate success (i.e. $f=0$). In this case there are clearly $n_l$ redundant arm qubits that must be removed by $X$-measurement. The second term sums over all the cases where at least one gate failure occurs. In such cases there with be $n_l-2f+1$ redundant qubits that must be removed by $X$-measurement, and one that must be removed by $Z$-measurement (the one preceding the successful {\sc CPHASE} gate). Note that in fact $f$ $Z$-measurements take place, but only the last one actually contributes to error accumulation (where we assume that when the {\sc CPHASE} gate fails it does not measure the qubits). In the case of a single-shot strategy, where a type-(b) cluster is prepared directly, $\langle m\rangle=0$. Next consider the single-shot success probability of preparing both the +- and (b)-type clusters. Clearly, we have
%\begin{eqnarray}
%p_\mathrm{type-a}&=&{p_\mathrm{gate}}^{4n_l}\nonumber\\
%p_\mathrm{type-b}&=&{p_\mathrm{gate}}^{8n_l+4}.
%\end{eqnarray}

A simple numerical example is illustrative. From Ref.~\cite{bib:Duan05}, constructing a 100 qubit cluster state with 10\% success probability, using {\sc CPHASE} gates that operate with 99\% success probability requires a resource of +-clusters with arm length $n_l\approx 11$. Suppose we construct the resource states using a single-shot approach to prevent accumulation of errors. Then the preparation of each resource +-cluster succeeds with probability $p_\mathrm{success}={p_\mathrm{gate}}^{4n_l}\approx 0.64$. The next step in the protocol is to join these +-clusters together to form clusters of type (b). We are in a high $p_\mathrm{gate}$ regime, so the average number of redundant qubits that must be measured away is $\approx n_l$. With a physical depolarizing rate of $p_\mathrm{error}=10^{-3}$, after measurement of the redundant qubits, this results in an effective depolarizing rate of $p_\mathrm{eff}=1.1\times 10^{-2}$, an order of magnitude increase. Alternately, we could produce the type-(b) clusters directly as a resource. In this case the single-shot success probability is $p_\mathrm{success}={p_\mathrm{gate}}^{6n_l+1}\approx 0.51$. However, now there are no accumulated errors associated with joining the +-clusters, so the effective error rate at this stage of the protocol is just the physical error rate of $10^{-3}$.

While this particular example exhibits a very significant reduction in the effective error rate, it is clear that we are limited to a high $p_\mathrm{gate}$ regime, as a result of the large exponents. Now the tradeoff becomes evident. For lower values of $p_\mathrm{gate}$, we loose our ability to directly prepare type-(b) clusters, and for yet lower values of $p_\mathrm{gate}$, to prepare +-clusters. Beyond this we have to resort to preparing the +-clusters non-deterministically, but we may still attempt to prepare the initial linear clusters via single-shot, and so on. While it is obvious that this approach is limited, the benefits of shifting as much of the state preparation into single-shot construction are clear.

\subsection{Discussion}
While we have analyzed a specific state preparation strategy, these results are likely to be applicable to other divide-and-conquer type strategies, which have applications beyond quantum computing. For example, Kieling \emph{et al.} \cite{bib:Kieling06} recently investigated optimal strategies for constructing cluster states using non-deterministic gates. Their analysis focussed on minimizing physical resource requirements, and was entirely classical. However, error blow-out is affected by more than just resource requirements. First, as we have demonstrated, it is highly dependent upon the preparation strategy. Second, as we know from our discussion in section \ref{sec:graph_state_disc}, it is also highly topology dependent. Thus, asking questions like ``which strategy minimizes the required number of gate operations?'' or ``which strategy minimizes physical resource requirements?'' overlook this important issue. We suggest that future investigations of state preparation strategies adopt a more rigorous definition of `optimal' which includes consideration of error propagation effects.

Let us qualify this statement further by considering two simple example state preparation strategies. First, consider a pure single-shot strategy. Clearly, when using probabilistic gates this approach has an exponentially small success probability and therefore requires exponentially large resources. This is one extreme of the example discussed above, where we shift the entire preparation into single-shot. Although this strategy has exponentially low success probability, it does not require the introduction of \emph{any} redundant qubits and therefore will not accumulate additional errors associated with unwanted qubits that must be measured out. Second, consider the other extreme, a divide-and-conquer approach, where we probabilistically build up a large cluster from numerous small clusters that are prepared offline in advance. The +-cluster approach discussed, as well as micro-cluster approaches, are examples of this. In general, this type of strategy exhibits polynomial resource requirements. However, while this strategy is superior from a resource perspective, it is inferior from an error propagation perspective, since it necessarily introduces redundant qubits which must be removed via measurement. In this simple comparison of two extreme cases we see that physical resource requirements and error accumulation are directly competing parameters.

\section{Loss-tolerant cluster states} \label{sec:cluster}
\begin{figure}[!htb]
\includegraphics[width=0.5\columnwidth]{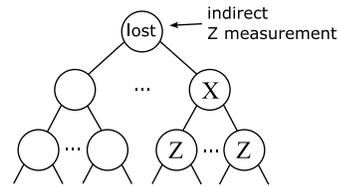}
\caption{Using a tree cluster to perform indirect $Z$-measurement of a lost qubit. An indirect $Z$-measurement is performed by measuring a qubit below the lost qubit in the $X$-eigenbasis, and each of the connected qubits below that in the $Z$-eigenbasis. If the $X$-measurement fails, we can make another attempt on the next branch. If any of the $Z$-measurements fail they can be indirectly measured by moving further down the tree.} \label{fig:cluster}
\end{figure}

We now examine the Varnava \emph{et al.} \cite{bib:Varnava05} scheme for loss-tolerant cluster state quantum computation. Here a cluster qubit is replaced with a `tree' cluster. This structure facilitates multiple attempts at \emph{indirect measurement} of the lost qubit, suppressing effective loss rates. Indirect measurement is a feature of cluster states that follows from the stabilizer representation. The stabilizers impose correlations in measurement outcomes of cluster state qubits. Indirect measurement exploits these correlations to infer the measurement result of a lost qubit using only the measurement results of correlated qubits, explained in Fig.~\ref{fig:cluster}.

Indirect measurement via the tree structure exhibits similar unfavorable error scaling characteristics to the previous example. Specifically, an indirect measurement outcome will be incorrect if an odd number of the involved measurements were incorrect. Thus, the effective error rate of the lost root qubit scales up exponentially with the number of measurements made. Furthermore, high loss tolerance is achieved by using larger trees. Therefore higher loss tolerance implies higher error rates, because more qubits are measured during the indirect measurement. A numerical example is illustrative. Based on results from Ref.~\cite{bib:Varnava05}, achieving an effective loss rate of $\varepsilon_\mathrm{eff}\approx 10^{-3}$ given a physical loss rate of $\varepsilon_\mathrm{loss}=0.2$, requires tree clusters with roughly $Q\approx 1000$ qubits. Suppose an indirect measurement requires measuring half the tree on average. This will magnify a physical error rate of $p_\mathrm{error}\approx 10^{-3}$ to an effective error rate on the indirectly measured qubit of $p_\mathrm{eff}\approx 0.32$, an increase of more than three orders of magnitude.

We now consider an approach for minimizing this effect, again relying on a majority voting technique. In fact, this modification does not require any changes to the tree structure, but only a change in the protocol. The tree is characterized by branching parameters $\{b_1,b_2,\dots,b_d\}$ (i.e. the number of branches emanating from each node of the respective level), where $d$ is the depth of the tree. To indirectly measure the lost root qubit we have as many attempts as there are branches at the first level of the tree, $b_1$. In the original protocol we simply keep attempting indirect measurement via the different branches until one succeeds. We now modify the protocol as follows. We simultaneously perform indirect measurement via \emph{all} available branches. From the ones which succeed we perform a majority vote to determine the correct indirect measurement outcome. If indirect measurement is performed in parallel via $b_1$ branches, the probability of an error propagating into the measurement outcome scales as $p_\mathrm{eff}=\textsc{Exp}^{-1}(b_1)$ with $p_\mathrm{single}$, the probability of any single indirect measurement being incorrect. On the other hand, $p_\mathrm{single}$ scales as $p_\mathrm{single}=\textsc{Exp}(\textsc{Poly}[b_2,\dots,b_d])$ with $p_\mathrm{error}$. Therefore, for an appropriate choice of branching parameters $\{b_i\}$, one expects that exponential error scaling can be eliminated.

We now analyze this modification in detail. For simplicity we begin by presenting the analysis for a simple two-level tree structure with uniform branching parameters. We then generalize the analysis to arbitrary tree structures.

\subsection{Two-level error propagation analysis}
We begin by considering a two-level tree structure with uniform branching parameters $b=b_1=b_2$. $p_\mathrm{local}$ and $p_\mathrm{loss}$ are the local error and loss rates respectively.

A single indirect measurement (i.e. along just one of the branches) succeeds if \emph{none} of the qubits involved were lost,
\begin{equation}
p_\mathrm{im-success}=(1-p_\mathrm{loss})^{b+1}.
\end{equation}
The effective loss rate is the probability that all $b$ indirect measurements fail,
\begin{equation}
p_\mathrm{loss}'=(1-p_\mathrm{im-success})^b.
\end{equation}
The probability of an error occurring during a single indirect measurement is the probability that an odd number of measurement errors occur on qubits involved in the indirect measurement,
\begin{equation}
p_\mathrm{im-error}=\sum_{i\in 2\mathbb{Z}+1}\binom{b}{i}{p_\mathrm{local}}^i(1-p_\mathrm{local})^{b-i}.
\end{equation}
% Note that $p_\mathrm{local}$ is the probability of a local measurement result being incorrect, which is half the depolarizing rate. Similarly, $p_\mathrm{im-error}$ must be interpreted as the probability of the indirect measurement result being incorrect, not the effective depolarizing rate.
The probability that $m$ indirect measurements succeed is
\begin{equation}
p_\mathrm{success}(m)=\binom{b}{m}{p_\mathrm{im-success}}^m(1-p_\mathrm{im-success})^{b-m}.
\end{equation}
The probability of an error being introduced after majority voting, given that $m$ indirect measurements succeed is
\begin{equation}
p_\mathrm{error}(m)=\sum_{i>m/2}{p_\mathrm{im-error}}^i(1-p_\mathrm{im-error})^{m-i}.
\end{equation}
The overall probability of an error being introduced after majority voting is
\begin{equation}
p_\mathrm{error}=\sum_{i=1}^b p_\mathrm{success}(i)p_\mathrm{error}(i).
\end{equation}

\subsection{General error propagation analysis}
The general analysis proceeds along the same lines as the two-level case. However, now the relationships are defined recursively. We use the notation $p^{(i)}$ to denote a probability at level $i$ of the tree structure.

Imagine we wish to measure the $Z$-observable of a qubit at level $i$. There are two ways in which this can succeed: either the photon was \emph{not} lost, and it can be measured directly, or it \emph{was} lost, but it is successfully measured indirectly. The probability that the later case succeeds is the probability that indirect measurement succeeds ($p_\mathrm{im-success}^{(i+1)}$) via \emph{any} of the underlying $b_i$ routes. Each of these indirect measurements will succeed if the qubit at level $i+1$ (which we measure in the $X$-eigenbasis) was \emph{not} lost, and $Z$-measurement on each of the qubits below that, at level $i+2$, succeed. Thus,
\begin{eqnarray}
p_\mathrm{Z-success}^{(i)}&=&(1-p_\mathrm{loss})+p_\mathrm{loss}\left[1-\left(1-p_\mathrm{im-success}^{(i)}\right)^{b_i}\right]\nonumber\\
p_\mathrm{im-success}^{(i)}&=&(1-p_\mathrm{loss}){p_\mathrm{Z-success}^{(i+2)}}^{b_{i+1}}
\end{eqnarray}
where $i$ is odd. The probability that $m$ indirect measurements succeed is
\begin{equation}
p_\mathrm{im-success}^{(i)}(m)=\binom{b_i}{m}{p_\mathrm{im-success}^{(i)}}^{m}\left(1-p_\mathrm{im-success}^{(i)}\right)^{b_i-m}.
\end{equation}
This reasoning applies for all levels $i$, except the top level ($i=1$) and bottom level ($i=d$). At the top level the root qubit has definitely been lost, so the expression reduces to
\begin{equation}
p_\mathrm{Z-success}^{(1)}=\left[1-\left(1-p_\mathrm{im-success}^{(i)}\right)^{b_i}\right].
\end{equation}
At the bottom level, indirect measurement is not possible, so $Z$-measurements will only succeed if the qubits have not been lost
\begin{equation}
p_\mathrm{Z-success}^{(d)}=1-p_\mathrm{loss}.
\end{equation}
Now let us turn our attention to error propagation. Again there are two possibilities. If the photon was \emph{not} lost, and measured directly, the probability of an error being picked up at that level is the local error rate. If the photon \emph{was} lost, an error will be picked up if the majority of the underlying indirect measurements suffered errors.
\begin{eqnarray}
p_\mathrm{maj-error}^{(i)}(m)&=&\sum_{j>m/2}p_\mathrm{im-error}^{(i)}(m)^{j}p_\mathrm{im-error}^{(i)}(m)^{m-j}\nonumber\\
p_\mathrm{maj-error}^{(i)}&=&\sum_{j=0}^{b_i}p_\mathrm{im-success}^{(i)}(m)p_\mathrm{maj-error}^{(i)}(m)\nonumber\\
p_\mathrm{error}^{(i)}&=&(1-p_\mathrm{loss})p_\mathrm{local}+p_\mathrm{loss}p_\mathrm{maj-error}^{(i)}.\nonumber\\
\end{eqnarray}
where $p_\mathrm{maj-error}^{(i)}(m)$ is the probability of the majority vote being incorrect given that $m$ indirect measurements are performed, $p_\mathrm{maj-error}^{(i)}$ is the net probability of the majority vote being incorrect, and $p_\mathrm{error}^{(i)}$ the total error probability. It is worth noting here that where indirect measurement is possible $p_\mathrm{maj-error}^{(i)}\leq p_\mathrm{local}$. This suggests a further optimization to our measurement strategy. Namely, even when a qubit is present, we should always preferentially measure through indirect measurement, rather than through direct measurement. This scenario can be modeled through a trivial modification of the previous equation.

Finally, the probability of an error being introduced during any given indirect measurement is the probability that an \emph{odd} number of measurement errors are introduced onto the involved qubits. There are two possibilities, either a local error occurs on the qubit being measured in the $X$-eigenbasis and an even number of errors occur on the qubits measured in the $Z$-eigenbasis, or no error occurs on the qubit being measured in the $X$-eigenbasis and an odd number of errors occur on the qubits measured in the $Z$-eigenbasis. Thus,
\begin{eqnarray}
p_\mathrm{im-error}^{(i)}=p_\mathrm{local}\sum_{i\in 2\mathbb{Z}}\binom{b_{i+1}}{i}{p_\mathrm{error}^{(i+2)}}^i\left(1-p_\mathrm{error}^{(i+2)}\right)^{b_{i+1}-i}\nonumber\\
+(1-p_\mathrm{local})\sum_{i\in 2\mathbb{Z}+1}\binom{b_{i+1}}{i}{p_\mathrm{error}^{(i+2)}}^i\left(1-p_\mathrm{error}^{(i+2)}\right)^{b_{i+1}-i}.\nonumber\\
\end{eqnarray}

\subsection{Discussion}
In the modified gate failure tolerant protocol we observed a tradeoff between failure and depolarizing rates. In this example we see a similar effect. In the modified protocol loss rates determine the \emph{effective} value of $b_1$. That is, the number of indirect measurements that succeed at the first level of the tree depends on the loss rate. Thus, higher loss rates imply lower confidence in the majority vote and therefore lower tolerance against depolarizing noise. This undermines the otherwise very high loss tolerance promised by this scheme and introduces a direct tradeoff between these two error types. Let us consider a simple numerical example to illustrate this point. We analyzed a simple two level tree-structure with branching parameters $b_1=b_2=3$. This structure improves the effective loss rate for $\varepsilon\lesssim 0.195$. That is, below this threshold the effective loss rate is lower than the physical loss rate. Using the original protocol, without majority voting, this loss rate would increase a physical error rate of $p_\mathrm{error}=10^{-3}$ to an effective error rate on the lost qubit of $p_\mathrm{eff}\approx 4\times 10^{-3}$. With the introduction of majority voting this reduces to $p_\mathrm{eff}\approx 1.7\times 10^{-3}$. More importantly, there is a `break-even' point on the physical loss rate, below which there is no degradation in the effective error rate (i.e. $p_\mathrm{eff}\leq p_\mathrm{error}$). In this example this occur at $\varepsilon\approx 0.1$, roughly half the in-principle loss tolerance rate. This leads to several conclusions. First, with the addition of majority voting this scheme is useful not only as a loss-tolerance technique, but also as a quantum error correction technique. Second, if we do not wish effective error rates to suffer, tolerable loss rates are significantly reduced.

\section{Loss tolerant parity states} \label{sec:parity}
\begin{figure}[!htb]
\includegraphics[width=0.6\columnwidth]{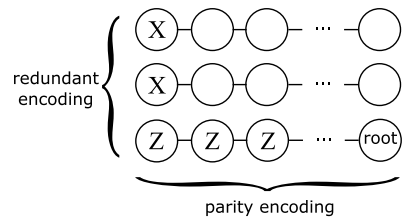}
\caption{Re-encoding in the parity state scheme. A root qubit is chosen from the initial encoded state. Every other qubit in the same level of the redundant encoding is measured in the $Z$-eigenbasis, and a single qubit from each of the other levels is measured in the $X$-eigenbasis. The remaining root qubit is used to construct a new encoded state.} \label{fig:parity}
\end{figure}

Finally we consider the Ralph \emph{et. al} \cite{bib:RalphHayes05} loss tolerant parity state scheme. In this scheme there are two levels of encoding. At the lower level, computational basis states are encoded as equal superpositions of odd or even parity states,
\begin{eqnarray}
\ket{0}_L&=&(\ket{+}^{\otimes n}+\ket{-}^{\otimes n})/\sqrt{2}\nonumber\\
\ket{1}_L&=&(\ket{+}^{\otimes n}-\ket{-}^{\otimes n})/\sqrt{2}
\end{eqnarray}
where $\ket\pm=(\ket{0}\pm\ket{1})/\sqrt{2}$, and $n$ is the level of parity encoding. These parity states are locally equivalent to maximally entangled GHZ states, and can therefore be regarded as $K_n$ graph states. Above the parity encoding is a level of $q$-fold redundant encoding,
\begin{equation}
\ket\psi_L=\alpha\ket{0}_L^{\otimes q}+\beta\ket{1}_L^{\otimes q}.
\end{equation}
One of the fundamental operations in this architecture is \emph{re-encoding}. Here new qubits are fused onto an existing parity state and the old ones measured out. This operation is used to `refresh' lost qubits, and is also required for the implementation of some quantum logic gates. The re-encoding procedure proceeds as follows. A `root' node is chosen, onto which a new redundant parity state is fused. All of the remaining qubits in the same level of redundant encoding as the root node are measured in the $Z$-eigenbasis. Additionally, a single qubit from every other level of the redundancy is measured in the $X$-eigenbasis. See Fig.~\ref{fig:parity}. Refer to Ref.~\cite{bib:RalphHayes05} for a detailed description.

Upon re-encoding, $N=n+q-1$ qubits must be measured. Let us consider how errors propagate in this context. As before, we will assume that physical qubits are all subject to independent depolarizing noise characterized by error probability $p$. Because the state is maximally entangled, upon measurement the residual state will be depolarized if \emph{any} of the $N$ measured qubits were depolarized. Thus, the effective error rate on the residual state is given by
\begin{equation}
p'=1-(1-p)^N.
\end{equation}
Again we see exponential dependence of the effective error rate on the number of qubits measured, and therefore the size of the parity state. In the same vein as the previous examples, this significantly undermines the loss tolerance of the scheme since practical error tolerance requirements effectively limit us to using small parity states.

We now briefly describe a technique for suppressing error propagation. We rely on the same technique as previously -- we introduce redundant qubits to facilitate majority voting. Due to the complexity of analyzing this scheme we do not provide a quantitative analysis. However it should be clear from the discussion that the described modification ought to exhibit similar features to the previous examples.

We modify the scheme by introducing a new layer of redundant encoding underneath the parity states. Specifically,
\begin{equation}
\ket{0}\to\ket{0}^{\otimes m},\ \ket{1}\to\ket{1}^{\otimes m}.
\end{equation}
This new layer of redundant encoding is introduced asymmetrically. The redundancy is applied to all but one qubit from each parity state. The non-redundant qubits are where fusions are attempted. Thus, under the modified protocol only one fusion attempt per parity state is possible. This requires increasing $q$ to provide sufficient fusion opportunities to suppress loss to the required level. When unwanted parity qubits are measured out in the $Z$-eigenbasis the new level of redundancy allows for majority voting on each measurement. Furthermore, when parity qubits are measured in the $X$-eigenbasis, rather than measure a \emph{single} qubit from the respective parity state (as per the original protocol), we measure \emph{all} qubits and perform a majority vote on the outcomes. Note that this only requires a change in the protocol, and does not require changes to the encoding. It is obvious that these modifications will suppress measurement errors. However, while this is effective at suppressing noise rates, it does so at the expense of loss-tolerance. When any parity qubit is measured in the $X$-eigenbasis, if \emph{any} of the underlying redundant qubits were lost, the qubit will be dephased and the measurement result randomized. Thus, for an $X$-measurement to be successful, \emph{all} underlying redundant qubits must be present. As per the previous examples, this presents us with a tradeoff between loss and error rates. Higher loss rates imply a lower success probability when performing $X$-measurements. Thus, when a parity state is measured in the $X$-eigenbasis, the confidence of the majority vote is reduced, thereby increasing effective noise rates.

\section{Conclusion} \label{sec:conclusion}
We have considered three well known protocols for tolerating gate failure and qubit loss. We demonstrated that while very effective at dealing with these particular error types, these schemes have the undesirable effect of magnifying other error types, namely depolarizing noise. In each case we discussed techniques for minimizing this effect. However, this introduces a tradeoff between loss/failure and depolarizing rates. In practical situations we always need some degree of tolerance against both these error types. This implies that in realistic scenarios these schemes are unlikely to achieve the loss/failure tolerance they are capable of in principle.

While we specifically considered three well-known protocols, we believe our results are likely to be applicable to other related protocols. For example, other state preparation strategies such as Nielsen's micro-cluster approach \cite{bib:Nielsen04} ought to exhibit similar error propagation properties, since they also rely on measuring out redundant qubits.

In conclusion, when designing gate failure or loss tolerant quantum computing protocols, it is extremely important to be mindful of error propagation characteristics. As we have demonstrated, specialized schemes which ignore these effects often boast misleadingly high loss and failure tolerance. While there are techniques for minimizing this problem, they significantly reduce the loss/failure tolerance of these schemes. Nonetheless, such schemes may be very useful for tolerating modest levels of loss and gate failure.

% ACKNOWLEDGMENTS
\begin{acknowledgments}
We thank Michael Nielsen for the discussion that motivated this work, and Henry Haselgrove and Alex Hayes for helpful discussions. The use of majority voting in tree clusters to suppress error rates was first recognized by Daniel Browne, Terry Rudolph and Michael Varnava. This work was supported by the Australian Research Council and QLD State Government. We acknowledge partial support by the DTO-funded U.S. Army Research Office Contract No. W911NF-05-0397
\end{acknowledgments}

% BIBLIOGRAPHY
\bibliography{paper}

\end{document}